\documentclass[12pt]{article}
\usepackage{amsfonts,amssymb,amsmath}
\usepackage[dvips]{epsfig}
\begin{document}
\title{\bf\large{The minimum-error discrimination via
Helstrom family of ensembles and Convex
Optimization}}\vspace{20mm}
\author{M. A. Jafarizadeh,$^{1,2,3}$\thanks{E-mail:jafarizadeh@tabrizu.ac.ir}\quad
Y. Mazhari,$^{1}$\thanks{E-mail:Mazhari@tabrizu.ac.ir}\quad and M. Aali$^{4}$\thanks{E-mail:s.aali@azaruniv.edu}\\
$^1${\small Department of Theoretical Physics and Astrophysics,
University of Tabriz, Tabriz 51664, Iran.}  \\ $^2${\small
Institute for Studies in Theoretical Physics and Mathematics,
Tehran 19395-1795, Iran.}\\$^3${\small Research Institute for
Fundamental Sciences, Tabriz 51664, Iran.}\\$^4${\small Department
of Physics, Faculty of Science, Azarbaijan University of Tarbiat
Moallem,}\\{\small 53714-161 Tabriz, Iran.}} \maketitle
\begin{abstract}
Using the \emph{convex optimization} method and \emph{Helstrom
family of ensembles}  introduced in Ref. \cite{Kimura}, we have
discussed optimal ambiguous discrimination in qubit systems. We
have analyzed the problem of the optimal discrimination of
N known quantum states and have obtained maximum success probability and
optimal measurement for N known quantum states with equiprobable prior
probabilities and equidistant from center of the Bloch ball, not
all of which are on the one half of the Bloch ball and all of the
\emph{conjugate states} are pure. An exact solution has also been
given for arbitrary three known quantum states. The given examples which use
our method include: 1. Diagonal N mixed states; 2. N equiprobable states
and equidistant from center of the Bloch ball which their
corresponding Bloch vectors are inclined at the equal angle from z
axis; 3. Three mirror-symmetric states; 4. States that have been
prepared with equal prior probabilities on vertices of a Platonic
solid.
\end{abstract}


\section{Introduction}
\par

In quantum information, the problem of detecting information stored
in the state of a quantum system is of fundamental interest. In
the simplest case, two dimensional systems or qubits can be used
to store quantum information. We assume that a quantum system is
prepared in a certain state that is drawn with known previous
probability from a finite set of known possible states and we want
to find the best possible measurement that can be used to
determine the actual state of the quantum system. If the states
are mutually orthogonal, then they can be distinguished perfectly.
But because of the quantum interference, it is impossible to
discriminate quantum states by measurement. There are two basic
approaches to accomplish state discrimination. In one approach,
which is called the \emph{minimum-error discrimination} and is
the ambiguous discrimination, measurement outcomes are not allowed
to be inconclusive but there is the possibility that the state is
identified incorrectly. In this case, the probability of
successful discrimination is made maximum by the optimum
measurement.
 In the other base approach, which is called
\emph{optimum unambiguous discrimination}, no error occurs, but
there exists a measurement outcome which gives an inconclusive
result where we fail to identify the state. In this approach, it is
tried to minimize the failure probability by appropriate
measurements. The topic of quantum state discrimination was firmly
established in the 1970s by the pioneering work of Helstrom
\cite{Holevo}, who considered a minimum error discrimination of
two known quantum states and unambiguous state discrimination was
originally formulated and analyzed by Ivanovic, Dieks, and Peres
\cite{Ivanovic,Dieks,Peres} in 1987. \cite{Jafari1} In this paper,
we deal with \emph{minimum-error discrimination} discrimination.
We will use \emph{convex optimization}  \cite{Boyd} as a tool for reaching this
aim, which has many other different applications where some of them have been seen
in previous papers
\cite{Jafari1,Jafari2,Jafari3,Jafari4,Jafari7,Mirzaee,Jafari5,Jafari6}.
\par
In the present work, we have used \emph{convex optimization} for discrimination of known quantum states with the aid of
\emph{Helstrom family of ensemble} idea in qubit systems. Applied
optimization problem is minimization of upper bound of optimal
success probability. Minimum upper bound is equal with maximum
success probability because there exist POVM elements that are
orthogonal to \emph{conjugate states} (by definition, any
\emph{conjugate state} have same convex combination with its
corresponding known state)
\cite{Kimura}. In this case, the
POVM elements give an optimal measurement to discriminate given
states. It has been proved, at least two \emph{conjugate states}
can be pure. If \emph{conjugate state} is pure, then its
corresponding optimal measurement is orthogonal to it and if
\emph{conjugate state} is mixed, then its corresponding optimal
measurement is zero operator. We have also shown, it is impossible
for all of \emph{Lagrange multipliers} associated with the
inequality constraints to take on the value zero. Optimal
measurement elements and optimal success probability have been
obtained for discriminating : 1. N equiprobable quantum states, which
distance between each of them and center of Bloch ball is equal
and, not all of which lie on the one half of Bloch ball, and all
of the \emph{conjugate states} are pure; 2. Arbitrary
three known quantum states; 3. Diagonal N mixed states; 4. N equiprobable quantum states
which distance between each of them and center of Bloch ball is
equal and their corresponding Bloch vectors are inclined at the
equal angle from z axis.
\par
The organization of this paper is as follows. In Sec. II, in summary we
illustrate the measurement operators of constituting POVM for
\emph{minimum-error discrimination} of quantum states and
\emph{Helstrom family of ensembles} as a strategy for carrying out
of discrimination. Then in Sec. III with transformation of the problem format
to optimization problem we obtain the KKT conditions \cite{Boyd}
and problem formulation, and then by using them we determine optimal
measurement operators and obtain optimal success probability for
N equiprobable quantum states located at equal distance from center of
Bloch ball, which all of the states are not on the one half of the
Bloch ball and all of the \emph{conjugate states} are on the
boundary of Bloch ball and then we solve the problem exactly for
two arbitrary quantum states, three arbitrary quantum states and some examples in Sec. IV which comprise: 1. Diagonal N mixed
states; 2. N quantum states with equal prior probability and equidistant
from center of the Bloch ball which their corresponding Bloch
vectors are inclined at the equal angle from z axis; 3. The
symmetrical mirror states $|\psi_{1}\rangle$,
$|\psi_{2}\rangle$ and $|\psi_{3}\rangle$ that subject to
transformation $|0\rangle\rightarrow|0\rangle$,
$|1\rangle\rightarrow-|1\rangle$ change into
$|\psi_{2}\rangle$, $|\psi_{1}\rangle$ and $|\psi_{3}\rangle$,
respectively; 4. Quantum states that have been prepared with equal prior
probabilities on vertices of a Platonic solid. Finally, a brief
conclusion and two appendices have been provided.
\par
\section{Ambiguous quantum state discrimination and Helstrom
family of ensembles}
\par
We assume that a quantum system
is prepared with some known prior probability, in some state
chosen from a finite collection of given known conceivable states.
We want to identify the actual state of the quantum system. A
state discriminating measurement determines probabilistically what
the actual state of the system belongs to the set of possible states.
We also assume, the state space $\large{\emph{s}}$ is a convex set in a
real vector space and an operator $e_{j}$ on $\large{\emph{s}}$ is
defined by an affine functional
$e_{j}(\rho_{i})=Tr(\rho_{i}M^{\dag}_{j}M_{j})$ from
$\large{\emph{s}}$ to $[0,1]$ that $M_{j}$ is quantum measurement
operator and $p(j|i)=Tr(\rho_{i}M^{\dag}_{j}M_{j})$ is the
probability to infer from the measurement that the system is in
the state $\rho_{j}$ if it has been prepared in a state
$\rho_{i}$. The state of the system after the measurement is
$\frac{M_{j}\rho_{i}M^{\dag}_{j}}{Tr(\rho_{i}M^{\dag}_{j}M_{j})}$.
The measurement operators satisfy the completeness equation
\begin{equation}
\sum^{N}_{j=1}M^{\dag}_{j}M_{j}=I.
\end{equation}
Defining $\Pi_{j}=M^{\dag}_{j}M_{j}$, then $\Pi_{j}$ is a
positive operator such that \cite{Nielsen}
\begin{equation}\label{completeness}
\sum^{N}_{j=1}\Pi_{j}=I
\end{equation}
and $p(j|i)=Tr(\rho_{i}\Pi_{j})$. Thus the set of operators
$\Pi_{j}$ are sufficient to determine the probabilities of the
different measurement outcomes. The operators $\Pi_{j}$ are known
as the POVM elements associated with the measurement that are
suitable for \emph{minimum-error discrimination}. The error
probability is expressed as
\begin{equation}
P_{err}=\sum^{N}_{i=1}\sum^{N}_{j=1,j\neq
i}p_{i}Tr(\rho_{i}\Pi_{j})
\end{equation}
Suppose we are given a state chosen from $\{\rho_{i}\}^{N}_{i=1}$
with a prior probability distribution
$\{p_{i}\}^{N}_{i=1}(p_{i}\geq 0,\sum_{i}p_{i}=1)$. Our goal is to
find an optimal measurement to maximize the success probability to
discriminate the states. It is sufficient to consider an N-valued
observable $\{\Pi_{i}\}^{N}_{i=1}$ from which we decide the state
was in $\rho_{i}$ when obtaining the output i. The success
probability is
\begin{equation}\label{Pro1}
P=1-P_{err}=\sum^{N}_{i=1}p_{i}Tr(\rho_{i}\Pi_{i}).
\end{equation}
The maximal success probability $P^{opt}$ is caused by the best
operators $\{\Pi_{i}\}^{N}_{i=1}$. In order to use the
\emph{minimum-error discrimination} strategy, we have to determine
the particular detection operators $\{\Pi_{i}\}^{N}_{i=1}$ that
maximize the right-hand side of the equation (\ref{Pro1}) under
the constraint (\ref{completeness}). We shall use a useful family
of ensembles which have been introduced in Ref. \cite{Kimura}
and is later shown to be closely related to an optimal state
discrimination strategy. A set of N-numbers
$\{\tilde{p}_{i},\rho_{i};1-\tilde{p}_{i},\tau_{i}\}^{N}_{i=1}$ is
called a \emph{weak Helstrom family (of ensembles)} if there exist
N-numbers of binary probability discriminations
$\{\tilde{p}_{i},1-\tilde{p}_{i}\}^{N}_{i=1}$ and states
$\{\tau_{i}\}^{N}_{i=1}$ satisfying
$\frac{p_{i}}{\tilde{p}_{i}}=p\leq 1$ and
\begin{equation}\label{helstrom}
\tilde{p}_{i}\rho_{i}+(1-\tilde{p}_{i})\tau_{i}=\tilde{p}_{j}\rho_{j}+(1-\tilde{p}_{j})\tau_{j}
\end{equation}
for any i,j=1,...,N.
\par
We assume that a priori probability distribution satisfies
$p_{i}\neq 0,1$ in order to remove trivial cases. $p$ and
$\tau_{i}$ are called  \emph{Helstrom ratio} and \emph{conjugate
state} to $\rho_{i}$, respectively. It has been proved that
\cite{Kimura}
\begin{equation}
P^{opt}\leq p.
\end{equation}
An observable $\{\Pi_{i}\}^{N}_{i=1}$ satisfies $P^{opt}=p$ if
$Tr(\tau_{i}\Pi_{i})=0$ for any $i=1,...,N$. In this case, the
observable $\{\Pi_{i}\}^{N}_{i=1}$ gives an optimal measurement to
discriminates $\{\rho_{i}\}^{N}_{i=1}$ and we call the family
$\{\tilde{p}_{i},\rho_{i};1-\tilde{p}_{i},\tau_{i}\}^{N}_{i=1}$
\emph{Helstrom family of ensembles} \cite{Kimura}.\\
With
\begin{equation}
\rho_{i}=\frac{1}{2}(I+\textbf{b}_{i}.\vec{\sigma}),\quad
\tau_{i}=\frac{1}{2}(I+\textbf{c}_{i}.\vec{\sigma})
\end{equation}
expression (\ref{helstrom}) can be written in terms of
$\textbf{c}_{i}$ and $\textbf{b}_{i}$ which are the corresponding
Bloch vectors to $\rho_{i}$ and $\tau_{i}$, respectively, as
\begin{equation}\label{helstrom2}
\tilde{p}_{i}\textbf{b}_{i}+(1-\tilde{p}_{i})\textbf{c}_{i}=\tilde{p}_{j}\textbf{b}_{j}+(1-\tilde{p}_{j})\textbf{c}_{j}.
\end{equation}
\section{Problem formulation}
\subsection{The case of N quantum states}
\par
In this paper, we have restricted ourselves to qubit systems. In
future, our method will be used in qutrit systems.
\par
We shall find optimal success probabilities and optimal
measurements for discrimination of states $\rho_{i}$, $i=1,...,N$
which have been prepared with prior probabilities $p_{i}$. We will
see that minimum \emph{Helstrom ratio} equals optimal success
probabilities.
\par
Our problem is
$$
\mathrm{to\quad minimize}\quad p,
$$
$$
\mathrm{subject\quad to}\quad|\textbf{c}_{i}|^{2}-1\leq0,\quad
i=1,...,N;
$$
$$
\tilde{p}_{1}\textbf{b}_{1}+(1-\tilde{p}_{1})\textbf{c}_{1}-\tilde{p}_{i}\textbf{b}_{i}-(1-\tilde{p}_{i})\textbf{c}_{i}=0,\quad
i=1,...,N$$ that have been formulated as an optimization problem (see
Appendix A). It follows that this problem has the Lagrangian
$$
L=p+\sum^{N}_{i=1}\lambda_{i}(x^{2}_{i}+y^{2}_{i}+z^{2}_{i}-1)
$$
$$
+\sum^{N-1}_{i=1}\nu_{3i-2}(\tilde{p}_{1}b_{1x}+(1-\tilde{p}_{1})x_{1}-\tilde{p}_{i+1}b_{(i+1)x}-(1-\tilde{p}_{i+1})x_{i+1})
$$
$$
+\sum^{N-1}_{i=1}\nu_{3i-1}(\tilde{p}_{1}b_{1y}+(1-\tilde{p}_{1})y_{1}-\tilde{p}_{i+1}b_{(i+1)y}-(1-\tilde{p}_{i+1})y_{i+1})
$$
\begin{equation}
+\sum^{N-1}_{i=1}\nu_{3i}(\tilde{p}_{1}b_{1z}+(1-\tilde{p}_{1})z_{1}-\tilde{p}_{i+1}b_{(i+1)z}-(1-\tilde{p}_{i+1})z_{i+1})
\end{equation}
where $\textbf{b}_{i}=(b_{ix},b_{iy},b_{iz})$ and
$\textbf{c}_{i}=(x_{i},y_{i},z_{i})$.
\par
The partial derivative of the Lagrangian with respect to $p$ and
$x_{i}$, $y_{i}$, $z_{i}$, $(1\leq i\leq N)$ must vanish. Thus, the KKT conditions
with respect to
$\vec{\nu}_{i}=(\nu_{3i-5},\nu_{3i-4},\nu_{3i-3})$, $(2\leq i\leq
N)$ are
\begin{equation}
|\textbf{c}_{i}|^{2}-1\leq0,\quad i=1,...,N;
\end{equation}
\begin{equation}\label{equality}
\tilde{p}_{1}\textbf{b}_{1}+(1-\tilde{p}_{1})\textbf{c}_{1}-\tilde{p}_{i}\textbf{b}_{i}-(1-\tilde{p}_{i})\textbf{c}_{i}=0,\quad
i=2,...,N;
\end{equation}
\begin{equation}
\lambda_{i}\geq 0,\quad i=1,...,N;
\end{equation}
\begin{equation}\label{derivative1}
1+\sum_{i=2}^{N}\vec{\nu}_{i}.(\textbf{c}_{1}-\textbf{c}_{i})=0;
\end{equation}
\begin{equation}\label{derivative2}
2\lambda_{1}\textbf{c}_{1}+(1-\tilde{p}_{1})\sum_{i=2}^{N}\vec{\nu}_{i}=0;
\end{equation}
\begin{equation}\label{derivative3}
2\lambda_{i}\textbf{c}_{i}-(1-\tilde{p}_{i})\vec{\nu}_{i}=0 ,\quad
i=2,...,N;
\end{equation}
\begin{equation}\label{slackness1}
\lambda_{i}(|\textbf{c}_{i}|^{2}-1)=0 ,\quad i=1,...,N.
\end{equation}
\par
By the relations (\ref{derivative1}), (\ref{derivative2}) and
(\ref{derivative3}) we can conclude that it is impossible which
$\lambda_{i}=0$,\quad$i=1,...,N$. Also, The KKT conditions
conclude there can be at least two of $\textbf{c}_{i}$, $1\leq
i\leq N$, so $|\textbf{c}_{i}|=1$.
\par
Now our aim will be to solve KKT conditions to find optimal
POVM elements. From (\ref{derivative1}), (\ref{derivative2}) and
(\ref{derivative3})
\begin{equation}\label{sum2}
\sum_{i=1}^{N}\frac{\lambda_{i}\textbf{c}_{i}}{1-\tilde{p}_{i}}=0
\end{equation}
and
\begin{equation}\label{result1}
\sum_{i=1}^{N}\frac{\lambda_{i}|\textbf{c}_{i}|^{2}}{1-\tilde{p}_{i}}=\frac{1}{2}
\end{equation}
and by calculating $\textbf{c}_{i}$ from (\ref{helstrom2}) and
then substituting it into (\ref{sum2}) we arrive
at
\begin{equation}\label{result5}
\textbf{c}_{j}=\frac{\sum_{i=1}^{N}\frac{\lambda_{i}(p_{i}\textbf{b}_{i}-p_{j}\textbf{b}_{j})}{(p-p_{i})^{2}}}{\sum_{i=1}^{N}\frac{\lambda_{i}(p-p_{j})}{(p-p_{i})^{2}}},\quad
j=1,...,N.
\end{equation}
\par
When $|\textbf{c}_{i}|=1$, we choose its corresponding measurement
operator orthogonal to $\tau_{i}$. Thus, with the aid of
(\ref{sum2}) and (\ref{result1}) the POVM elements are found as
\begin{equation}\label{elem}
\Pi_{j}=\frac{4p\lambda_{j}}{p-p_{j}}|\chi_{j}\rangle\langle\chi_{j}|,\quad
j=1,...,N
\end{equation}
where
\begin{equation}\label{measur}
|\chi_{j}\rangle\langle\chi_{j}|=\frac{1}{2}(I-\textbf{c}_{j}.\vec{\sigma})
\end{equation}
and when $|\textbf{c}_{i}|<1$, the state $\tau_{i}$ is mixed and
$\Pi_{i}$ corresponding to $\textbf{c}_{i}$ is considered zero
operator (and therefore $\lambda_{i}=0$) in order that
$Tr(\tau_{i}\Pi_{i})=0$ is satisfied for all $i=1,...,N$ and
condition of $P^{opt}=p$ is provided. The terms corresponding to
all states $\rho_{i}$'s that \emph{conjugate states} to them are
mixed states do not have contributions to the sum in the relation
(\ref{Pro1}).
\par
If states have been prepared with equal prior
probabilities, then (\ref{sum2}), (\ref{result1}) and
(\ref{result5}) become
\begin{equation}\label{sum1}
\sum_{i=1}^{N}\lambda_{i}\textbf{c}_{i}=0;
\end{equation}
\begin{equation}
\sum^{N}_{i=1}\lambda_{i}|\textbf{c}_{i}|^{2}=\frac{Np-1}{2Np};
\end{equation}
\begin{equation}\label{blo1}
\textbf{c}_{j}=\frac{\textbf{D}}{(Np-1)\sum^{N}_{i=1}\lambda_{i}}-\frac{\textbf{b}_{j}}{Np-1},\quad
j=1,...,N,
\end{equation}
respectively, that we have defined
$\textbf{D}=\sum^{N}_{i=1}\lambda_{i}\textbf{b}_{i}$. Therefore,
\begin{equation}\label{blo3}
|\textbf{c}_{k}|^{2}-|\textbf{c}_{j}|^{2}=\frac{\textbf{b}^{2}_{k}-\textbf{b}^{2}_{j}}{(Np-1)^{2}}+\frac{2\textbf{D}.(\textbf{b}_{j}-\textbf{b}_{k})}{(Np-1)^{2}\sum^{N}_{i=1}\lambda_{i}},\quad
j,k=1,...,N.
\end{equation}
\par
As a special case we suppose, all of the states
$\rho_{1}$,...,$\rho_{N}$ are not on the one half of the Bloch
ball $(N\geq4)$ and their corresponding Bloch vectors have equal
length of $b$. We also suppose $|\textbf{c}_{i}|=1$ for all of the
vectors $\textbf{c}_{i}$, $i=1,...,N$ and then we can result
$\textbf{D}=0$. Therefore,
\begin{equation}\label{blo2}
|\chi_{j}\rangle\langle\chi_{j}|=\frac{1}{2}(I+\frac{\textbf{b}_{j}.\vec{\sigma}}{Np-1}),\quad
j=1,...,N
\end{equation}
and
\begin{equation}\label{pro4}
P^{opt}=p=\frac{1}{N}(1+b)
\end{equation}
where we have used $|\textbf{c}_{i}|=1$ for some $i$.
\par
In the next two subsections, we precisely work out the maximum
success probability and the optimal POVM elements for ambiguously
discriminating between any two-states, with prior probabilities
$p_{1}$, $p_{2}$ and among any three-states, with prior
probabilities $p_{1}$, $p_{2}$, $p_{3}$.
\subsection{The case of two quantum states}
\par
Although the case of two qubit states is studied \cite{Helstrom,Barnett}  we find instructive to see how the known solution follows from our method.\\
Equations (\ref{derivative2}), (\ref{derivative3}) and
(\ref{result1}) in this case are simply
\begin{equation}\label{derivative5}
2\lambda_{1} \textbf{c}_{1}+(1-\tilde{p}_{1})\vec{\nu}_{2}=0,
\end{equation}
\begin{equation}\label{derivative6}
2\lambda_{2} \textbf{c}_{2}-(1-\tilde{p}_{2})\vec{\nu}_{2}=0
\end{equation}
and
\begin{equation}\label{result2}
(1-\tilde{p}_{1})(1-\tilde{p}_{2})-2\lambda_{1}(1-\tilde{p}_{2})-2\lambda_{2}(1-\tilde{p}_{1})=0,
\end{equation}
respectively.
\par
The considerations $|\textbf{c}_{1}|=1$, $|\textbf{c}_{2}|=1$ and
equations (\ref{derivative5}), (\ref{derivative6}) and
(\ref{result2}) can be used to drive
\begin{equation}\label{lambda}
\lambda_{1}=\frac{1-\tilde{p}_{1}}{4},\quad
\lambda_{2}=\frac{1-\tilde{p}_{2}}{4}
\end{equation}
and substituting these values into equation (\ref{result5}) gives
\begin{equation}
\textbf{c}_{1}=\frac{p_{2}\textbf{b}_{2}-p_{1}\textbf{b}_{1}}{2p-1},\quad
\textbf{c}_{2}=-\textbf{c}_{1}.
\end{equation}
Using $|\textbf{c}_{1}|=1$ and paying attention to $p\geq p_{1}$
and $p\geq p_{2}$ we obtain
\begin{equation}
P^{opt}=p=\frac{1}{2}(1+|p_{2}\textbf{b}_{2}-p_{1}\textbf{b}_{1}|).
\end{equation}
Note that
\begin{equation}
\Pi_{1}=\frac{1}{2}(I-\frac{p_{2}\textbf{b}_{2}-p_{1}\textbf{b}_{1}}{2p-1}.\vec{\sigma}),\quad
\Pi_{2}=\frac{1}{2}(I+\frac{p_{2}\textbf{b}_{2}-p_{1}\textbf{b}_{1}}{2p-1}.\vec{\sigma}).
\end{equation}
\par
The minimum error probability $p^{min}_{err}$ is found to be
\begin{equation}
p^{min}_{err}=\frac{1}{2}(1-|p_{2}\textbf{b}_{2}-p_{1}\textbf{b}_{1}|).
\end{equation}
It can be written as
\begin{equation}
p^{min}_{err}=\frac{1}{2}(1-Tr|p_{2}\rho_{2}-p_{1}\rho_{1}|)
\end{equation}
which was originally found by Helstrom \cite{Helstrom}
where $\frac{1}{2}Tr|p_{2}\rho_{2}-p_{1}\rho_{1}|$ is the trace distance.\\
It is obvious that, the minimum error probability is achieved
when $\Pi_{1}$ and $\Pi_{2}$ are the projectors onto
eigenstates of $p_{2}\rho_{2}-p_{1}\rho_{1}$ that belong to
eigenvalues
$\frac{p_{2}-p_{1}}{2}-\frac{|p_{2}\textbf{b}_{2}-p_{1}\textbf{b}_{1}|}{2}$
and
$\frac{p_{2}-p_{1}}{2}+\frac{|p_{2}\textbf{b}_{2}-p_{1}\textbf{b}_{1}|}{2}$,
respectively \cite{Herzog}.
\subsection{The case of three quantum states }
\par
We now want to obtain the exact solution for discrimination of
three arbitrary known mixed states. We place N=3 on the equations
(\ref{derivative2}), (\ref{derivative3}) and (\ref{result1}) then
we have
\begin{equation}\label{derivative8}
2\lambda_{1}\textbf{\textbf{c}}_{1}+(1-\tilde{p}_{1})(\vec{\nu}_{2}+\vec{\nu}_{3})=0,
\end{equation}
\begin{equation}\label{derivative9}
2\lambda_{2}\textbf{c}_{2}-(1-\tilde{p}_{2})\vec{\nu}_{2}=0,
\end{equation}
\begin{equation}\label{derivative30}
2\lambda_{3}\textbf{c}_{3}-(1-\tilde{p}_{3})\vec{\nu}_{3}=0
\end{equation}
and
\begin{equation}\label{derivative11}
(1-\tilde{p}_{1})(1-\tilde{p}_{2})(1-\tilde{p}_{3})-2\lambda_{1}(1-\tilde{p}_{2})(1-\tilde{p}_{3})|\textbf{c}_{1}|^{2}-2\lambda_{2}(1-\tilde{p}_{1})(1-\tilde{p}_{3})|\textbf{c}_{2}|^{2}-2\lambda_{3}(1-\tilde{p}_{1})(1-\tilde{p}_{2})|\textbf{c}_{3}|^{2}=0.
\end{equation}
\par
If now we make the assumption $|\textbf{c}_{1}|=1$,
$|\textbf{c}_{2}|=1$, $|\textbf{c}_{3}|\neq1$ then
(\ref{derivative30}) becomes $\vec{\nu}_{3}=0$ and
(\ref{derivative8}), (\ref{derivative9}) and (\ref{derivative11}) are led back to
(\ref{derivative5}), (\ref{derivative6}) and (\ref{result2}),
respectively. In analogy to the two states case, the results are
given by
\begin{equation}
\lambda_{1}=\frac{1-\tilde{p}_{1}}{4},\quad\lambda_{2}=\frac{1-\tilde{p}_{2}}{4},\quad
\lambda_{3}=0;
\end{equation}
\begin{equation}
\textbf{c}_{1}=\frac{p_{2}\textbf{b}_{2}-p_{1}\textbf{b}_{1}}{2p-p_{1}-p_{2}}
,\quad\textbf{c}_{2}=-\textbf{c}_{1},\quad\textbf{c}_{3}=\frac{p_{1}\textbf{b}_{1}-p_{3}\textbf{b}_{3}}{p-p_{3}}+\frac{(p-p_{1})(p_{2}\textbf{b}_{2}-p_{1}\textbf{b}_{1})}{(p-p_{3})(2p-p_{1}-p_{2})};
\end{equation}
\begin{equation}\label{pro2}
P^{opt}=p=\frac{1}{2}(p_{1}+p_{2}+|p_{2}\textbf{b}_{2}-p_{1}\textbf{b}_{1}|);
\end{equation}
\begin{equation}
\Pi_{1}=\frac{1}{2}(I-\frac{p_{2}\textbf{b}_{2}-p_{1}\textbf{b}_{1}}{2p-p_{1}-p_{2}}.\vec{\sigma}),\quad
\Pi_{2}=\frac{1}{2}(I+\frac{p_{2}\textbf{b}_{2}-p_{1}\textbf{b}_{1}}{2p-p_{1}-p_{2}}.\vec{\sigma}),\quad
\Pi_{3}=0.
\end{equation}
The cases of $|\textbf{c}_{1}|\neq1$, $|\textbf{c}_{2}|=1$, $|\textbf{c}_{3}|=1$ and $|\textbf{c}_{1}|=1$, $|\textbf{c}_{2}|\neq1$, $|\textbf{c}_{3}|=1$ have similar results for optimal
success probability and the optimal POVM elements.
\par
Now we consider $|\textbf{c}_{1}|=1$, $|\textbf{c}_{2}|=1$,
$|\textbf{c}_{3}|=1$. Using
$[\tilde{p}_{i}\textbf{b}_{i}-\tilde{p}_{j}\textbf{b}_{j}]^{2}=[(1-\tilde{p}_{j})\textbf{c}_{j}-(1-\tilde{p}_{i})\textbf{c}_{i}]^{2}$,
we arrive at
\begin{equation}\label{dot1}
\textbf{c}_{i}.\textbf{c}_{j}=\frac{(p-p_{i})^{2}+(p-p_{j})^{2}-(p_{i}\textbf{b}_{i}-p_{j}\textbf{b}_{j})^{2}}{2(p-p_{i})(p-p_{j})},\quad
i,j=1,2,3.
\end{equation}
Substituting (\ref{dot1}) for $\textbf{c}_{1}.\textbf{c}_{2}$,
$\textbf{c}_{1}.\textbf{c}_{3}$ and
$\textbf{c}_{2}.\textbf{c}_{3}$ into (see Appendix B)
\begin{equation}\label{result3}
(\textbf{c}_{1}.\textbf{c}_{2})^{2}+(\textbf{c}_{1}.\textbf{c}_{3})^{2}+(\textbf{c}_{2}.\textbf{c}_{3})^{2}=2(\textbf{c}_{1}.\textbf{c}_{2})(\textbf{c}_{1}.\textbf{c}_{3})(\textbf{c}_{2}.\textbf{c}_{3})+1,
\end{equation}
we obtain
\begin{equation}
P^{opt}=p=\frac{-M+\sqrt{M^{2}-4LN}}{2L}
\end{equation}
that
$$
L=4(-p_{1}p_{2}+p_{1}p_{3}+p_{2}p_{3}-p^{2}_{3})I+4(p_{1}p_{2}-p_{1}p_{3}+p_{2}p_{3}-p^{2}_{2})J+4(p_{1}p_{2}+p_{1}p_{3}-p_{2}p_{3}-p^{2}_{1})K
$$
$$
+2IJ+2IK+2JK-I^{2}-J^{2}-K^{2},
$$
$$
M=2[(p_{1}^{2}p_{2}+p_{1}p_{2}^{2}-p_{1}^{2}p_{3}-p_{1}p_{3}^{2}-p_{2}^{2}p_{3}-p_{2}p_{3}^{2}+2p_{3}^{3})-p_{1}K-p_{2}J+p_{3}I]I
$$
$$
+2[(-p_{1}^{2}p_{2}-p_{1}p_{2}^{2}+p_{1}^{2}p_{3}+p_{1}p_{3}^{2}-p_{2}^{2}p_{3}-p_{2}p_{3}^{2}+2p_{2}^{3})-p_{1}K+p_{2}J-p_{3}I]J
$$
$$
+2[(-p_{1}^{2}p_{2}-p_{1}p_{2}^{2}-p_{1}^{2}p_{3}-p_{1}p_{3}^{2}+p_{2}^{2}p_{3}+p_{2}p_{3}^{2}+2p_{1}^{3})+p_{1}K-p_{2}J-p_{3}I]K,
$$
$$
N=[-p_{1}^{2}p_{2}^{2}+p_{1}^{2}p_{3}^{2}+p_{2}^{2}p_{3}^{2}-p_{3}^{4}+p_{1}^{2}K+p_{2}^{2}J-p_{3}^{2}I]I
$$
$$
+[p_{1}^{2}p_{2}^{2}-p_{1}^{2}p_{3}^{2}+p_{2}^{2}p_{3}^{2}-p_{2}^{4}+p_{1}^{2}K-p_{2}^{2}J+p_{3}^{2}I]J
$$
$$
+[p_{1}^{2}p_{2}^{2}+p_{1}^{2}p_{3}^{2}-p_{2}^{2}p_{3}^{2}-p_{1}^{4}-p_{1}^{2}K+p_{2}^{2}J+p_{3}^{2}I]K-IJK,
$$
$$
I=(p_{1}\textbf{b}_{1}-p_{2}\textbf{b}_{2})^{2},\quad
J=(p_{1}\textbf{b}_{1}-p_{3}\textbf{b}_{3})^{2},\quad
K=(p_{2}\textbf{b}_{2}-p_{3}\textbf{b}_{3})^{2}.
$$
Furthermore, by substituting $\textbf{c}_{1}.\textbf{c}_{2}$,
$\textbf{c}_{1}.\textbf{c}_{3}$ and
$\textbf{c}_{2}.\textbf{c}_{3}$ from (\ref{dot1}) into
(\ref{lambda1}), (\ref{lambda2}) and (\ref{lambda3}), we can
therefore give $\lambda_{1}$, $\lambda_{2}$ and $\lambda_{3}$ in
terms of the known quantities. Thus required optimal POVM
elements are determined by means of (\ref{result5}), (\ref{elem})
and (\ref{measur}).
\section{Examples}
\par
\textbf{1.} We will consider the situation in which given states
are
\begin{equation}
\rho_{i}=\frac{1}{2}(I+b_{iz}\sigma_{z})=\frac{1}{2}[(1+b_{iz})|0\rangle\langle0|+(1-b_{iz})|1\rangle\langle1|],\quad
i=1,...,N.
\end{equation}
Then relation (\ref{result5}) follows that the x- and
y- components of vectors $\textbf{c}_{i}$, $i=1,...,N$ are zero,
and thus there are only two vectors of $\textbf{c}_{i}$,
$i=1,...,N$ which have length $1$. While the last $N-2$ of POVM
elements can be chosen zero operator, the relation (\ref{sum2})
leads to the requirement
\begin{equation}\label{equal}
\frac{\lambda_{1}}{p-p_{1}}=\frac{\lambda_{2}}{p-p_{2}}.
\end{equation}
Using the z-component of the condition (\ref{helstrom2}) and
combining the relations (\ref{equal}), (\ref{result1}) and (\ref{elem}), we can
obtain the following optimal results.
\begin{equation}
\Pi_{1}=|0\rangle\langle0|,\quad\Pi_{2}=|1\rangle\langle1|\quad\mathrm{if}
\quad p_{1}b_{1z}\leq p_{2}b_{2z}
\end{equation}
and
\begin{equation}
\Pi_{1}=|1\rangle\langle1|,\quad\Pi_{2}=|0\rangle\langle0|\quad\mathrm{if}
\quad p_{1}b_{1z}\geq p_{2}b_{2z}
\end{equation}
and
\begin{equation}\label{probabi}
P^{opt}=\frac{1}{2}(p_{1}+p_{2}+|p_{2}b_{z2}-p_{1}b_{z1}|),
\end{equation}
where $b_{1z}$ and $b_{2z}$ are components that maximize the right
side of the relation (\ref{probabi}) over all $b_{iz}$.
\par
Similar to the two state case, the operators $\Pi_{1}$ and
$\Pi_{2}$ are the projectors onto
eigenstates of $p_{2}\rho_{2}-p_{1}\rho_{1}$.\\

\textbf{2.} Let us consider N mixed states $\rho_{j}$, $j=1,...,N$
with the corresponding Bloch vectors as
\begin{equation}
\textbf{b}_{j}=(b\sin\theta\cos\varphi_{j},b\sin\theta\sin\varphi_{j},b\cos\theta),\quad
j=1,...,N
\end{equation}
with uniform prior probability distribution ($p_{j}=\frac{1}{N}$,
$ j=1,...,N$). By (\ref{sum1}) we see
\begin{equation}
\sum_{j=1}^{N}\lambda_{j}z_{j}=0,
\end{equation}
which implies that vectors $\textbf{c}_{1},...,\textbf{c}_{N}$ are
in the $z=0$ plane. Furthermore relation (\ref{equality})
concludes that $\textbf{b}_{j}-\textbf{b}_{i}$ is parallel with
$\textbf{c}_{i}-\textbf{c}_{j}$ and
\begin{equation}
p=\frac{1}{N}+\frac{|\textbf{b}_{j}-\textbf{b}_{i}|}{N|\textbf{c}_{j}-\textbf{c}_{i}|},\quad i\neq j.
\end{equation}
Therefore $p$ is minimum if $|\textbf{c}_{i}|=1$, $i=1,...,N$.
\par
Equation (\ref{blo3}) concludes, $\textbf{D}$ is orthogonal to all
$\textbf{b}_{i}-\textbf{b}_{j}$, $i,j=1,...,N$, $i\neq j$. Thus
\begin{equation}
\textbf{D}=(0,0,b\cos\theta\sum^{N}_{i=1}\lambda_{i}).
\end{equation}
By applying (\ref{blo1}), we can show
\begin{equation}
\textbf{c}_{j}=\frac{b\cos\theta\hat{k}-\textbf{b}_{j}}{Np-1}=-\frac{b\sin\theta}{Np-1}(\cos\varphi_{j},\sin\varphi_{j},0)
\end{equation}
and note that $|\textbf{c}_{i}|=1$ and $p\geq p_{i},\quad
i=1,...,N$, we can obtain
\begin{equation}\label{probability1}
p=\frac{1}{N}(1+b\sin\theta).
\end{equation}
The corresponding \emph{conjugate states} to $\rho_{j}$,
$j=1,...,N$ are
\begin{equation}
|\phi_{j}\rangle=\cos(\frac{\pi}{4})|0\rangle+\sin(\frac{\pi}{4})e^{i(\varphi_{j}+\pi)}|1\rangle
,\quad j=1,...,N.
\end{equation}
By substituting (\ref{probability1}) in equation (\ref{elem})
we obtain
\begin{equation}
\Pi_{j}=\frac{4\lambda_{j}(1+b\sin\theta)}{b\sin\theta}|\chi_{j}\rangle\langle\chi_{j}|,\quad
j=1,...,N.
\end{equation}
We must have
\begin{equation}
|\chi_{j}\rangle=\cos(\frac{\pi}{4})|0\rangle+\sin(\frac{\pi}{4})e^{i\varphi_{j}}|1\rangle
\end{equation}
because $Tr(\tau_{j}|\chi_{j}\rangle\langle\chi_{j}|)=0$ must be satisfied
for all $j=1,...,N$. Hence $P^{opt}=p$ and
$|\chi_{j}\rangle\langle\chi_{j}|$, $j=1,..,N$ are optimal
measurement operator for discriminating the states $\rho_{j}$,
$j=1,...,N$.
\par
These results accord with given results in example 4 of Ref.
\cite{Kimura}.\\

\textbf{3.} For now we will consider the example in which the
input states are given by
\begin{equation}
|\psi_{1}\rangle=\cos\theta|0\rangle+\sin\theta|1\rangle
\end{equation}
\begin{equation}
|\psi_{2}\rangle=\cos\theta|0\rangle-\sin\theta|1\rangle
\end{equation}
\begin{equation}
|\psi_{3}\rangle=|0\rangle,
\end{equation}
with prior probabilities $p_{1}, p_{2}(=p_{1})$ and
$p_{3}=1-2p_{1}$. On the one hand, if $|\textbf{c}_{1}|=1$,
$|\textbf{c}_{2}|=1$ and $|\textbf{c}_{3}|\neq1$ then the relation
(\ref{pro2}) gives
\begin{equation}\label{pro5}
P^{opt}=p_{1}(1+\sin2\theta),
\end{equation}
on the other hand if $|\textbf{c}_{1}|=1$,
$|\textbf{c}_{2}|=1$ and $|\textbf{c}_{3}|=1$, we can use
(\ref{dot1}) to show
\begin{equation}\label{dot2}
\textbf{c}_{1}.\textbf{c}_{2}=\frac{(p-p_{1})^{2}-p^{2}_{1}(1-\cos4\theta)}{(p-p_{1})^{2}},
\end{equation}
\begin{equation}\label{dot3}
\textbf{c}_{1}.\textbf{c}_{3}=\textbf{c}_{2}.\textbf{c}_{3}=\frac{p^{2}-p_{1}p-p+p_{1}(1-2p_{1})\cos2\theta}{(p-p_{1})(p+2p_{1}-1)}
\end{equation}
and (\ref{result3}) evidently leads to
\begin{equation}\label{result4}
(\textbf{c}_{1}.\textbf{c}_{3})^{2}=\frac{1+\textbf{c}_{1}.\textbf{c}_{2}}{2}
\end{equation}
and by substituting (\ref{dot2}) and (\ref{dot3}) in (\ref{result4}),
the optimal success probability can be attained. It
is
\begin{equation}\label{pro6}
P^{opt}=\frac{(1-2p_{1})(p_{1}\sin^{2}\theta+1-2p_{1}-p_{1}\cos^{2}\theta)}{1-2p_{1}-p_{1}\cos^{2}\theta}.
\end{equation}
Therefore, there are two regimes depending on $p_{1}$ and $\theta$
which coincide at
$p^{'}_{1}=\frac{1}{2+\cos\theta(\sin\theta+\cos\theta)}$. If
$p_{1}\geq p^{'}_{1}$ ($p_{1}\leq p^{'}_{1}$), the optimal success
probability is given by the relation (\ref{pro6})(the relation
(\ref{result4})) \cite{Andersson}.
\par
In Ref. \cite{Andersson}, the measurement strategy of
\emph{minimum-error discrimination} has been used for obtaining the
relations (\ref{pro5}) and (\ref{pro6}).\\

\textbf{4.} In this example, we consider states $\rho_{i}$,
$i=1,...,N$ with equal prior probabilities, which form the
vertices of a Platonic solid centered at the origin with length of
edge $a$. By using relations (\ref{blo2}) and (\ref{pro4}), we
have given optimal success
probability  and the Bloch vectors of conjugate states.\\
Pyramid:
\begin{equation}
\textbf{c}_{i}=-\frac{\textbf{b}_{i}}{4P^{opt}-1},\quad i=1,...,4;\quad P^{opt}=\frac{1}{4}(1+\sqrt{\frac{3}{8}}a)
\end{equation}
Cube:
\begin{equation}
\textbf{c}_{i}=-\frac{\textbf{b}_{i}}{8P^{opt}-1},\quad i=1,...,8;\quad P^{opt}=\frac{1}{8}(1+\frac{\sqrt{3}}{2}a)
\end{equation}
Octahedron:
\begin{equation}
\textbf{c}_{i}=-\frac{\textbf{b}_{i}}{6P^{opt}-1},\quad i=1,...,6;\quad P^{opt}=\frac{1}{6}(1+\frac{\sqrt{2}}{2}a)
\end{equation}
Dodecahedron:
\begin{equation}
\textbf{c}_{i}=-\frac{\textbf{b}_{i}}{20P^{opt}-1},\quad
i=1,...,20;\quad P^{opt}=\frac{1}{20}(1+\frac{1}{3}a)
\end{equation}
Icosahedron:
\begin{equation}
\textbf{c}_{i}=-\frac{\textbf{b}_{i}}{12P^{opt}-1},\quad
i=1,...,12;\quad P^{opt}=\frac{1}{12}(1+\frac{\sqrt{5+\sqrt{5}}}{2\sqrt{2}}a)
\end{equation}
\section{Conclusion}
\par
Using the idea of \emph{Helstrom family}, there is one method for
ambiguous discrimination. In this method, \emph{Helstrom ratio} is considered to be the \emph{cost function}, subject to resulted
constraints of \emph{Helstrom family of ensembles}. If KKT
conditions associated with the optimization problem are satisfied,
minimum \emph{Helstrom ratio} will be equal with maximum success
probability. At least, two \emph{conjugate states} to known states
can be pure. Every optimal non-zero POVM element is orthogonal to
its corresponding pure conjugate state and all of the optimal POVM
elements corresponding to all mixed conjugate states are zero
operators. It is not possible for all of \emph{Lagrange
multipliers} associated with the inequality constraints to take on
the value zero. Our method has been restricted in qubit systems.
\par
Form of state space is not a sphere for qutrit systems and the located
states in boundary of state space are not necessarily pure,
therefore using this way for qutrit systems seems different and it
will be investigated in the future.\\

\vspace{1cm}\setcounter{section}{0} \setcounter{equation}{0}
\renewcommand{\theequation}{A-\roman{equation}}
{\Large{Appendix A}}\\

\textbf{A summary of convex optimization}\\
\par
An optimization problem \cite{Boyd} has the standard form:
$$
\mathrm{minimize}\quad f_{0}(x),
$$
$$
\mathrm{subject\quad to}\quad f_{i}(x)\leqslant 0,\quad i=1,...,m,
$$
\begin{equation}\label{opt}
h_{i}(x)= 0,\quad i=1,...,p,
\end{equation}
where the vector $x=(x_{1},...,x_{n})$ is called the
\emph{optimization variable} and the function $f_{0} :
\textbf{R}^{n} \rightarrow \textbf{R}$ the \emph{cost function}.
The inequalities $f_{i}(x)\leq 0$ are called \emph{inequality
constraints}, and the equations $h_{i}(x)=0$ are called the
\emph{equality constraints}.
\par
The \emph{Lagrangian L} : $\textbf{R}^{n}\times \textbf{R}^{m}
\times\textbf{R}^{p}\rightarrow \textbf{R}$ is
\begin{equation}\label{lag}
    L(x, \lambda, \nu)=f_{0}(x) + \sum_{i=1}^{m}\lambda_{i} f_{i}(x)+\sum_{i=1}^{p}\nu_{i}
    h_{i}(x).
\end{equation}
We refer to $\lambda_{i}$ as the \emph{Lagrange multiplier}
associated with the ith inequality constraint $f_{i}(x)\geq 0$;
similarly we refer to $\nu_{i}$ as the Lagrange multiplier
associated with the ith equality constraint $h_{i}(x)$. The
vectors $\lambda$ and $\nu$ are $(\lambda_{1},...,\lambda_{m})$
and $(\nu_{1},...,\nu_{p})$, respectively.
\par
The \emph{dual function} $g:\textbf{R}^{m+p}\rightarrow \textbf{R}$ is
defined as the minimum value of the Lagrangian over x that is
written as
\begin{equation}\label{duf}
    g(\lambda, \nu)=inf_{{x\in D}}L(x, \lambda, \nu)
\end{equation}
The dual function yields lower bounds on the optimal value
$p^{\star}$ of the problem (\ref{opt}): for any $\lambda\succeq0$
and any $\nu$ we have
\begin{equation}\label{lowerBound}
g(\lambda, \nu)\leqslant p^{\star}.
\end{equation}
\par
A natural question is: what is the
best lower bound that can be obtained from the Lagrange dual function?
This leads to the optimization problem\\
$$
\mathrm{maximize}\quad g(\lambda,\nu)\
$$
\begin{equation}\label{dual}
\mathrm{subject\quad to}\quad\lambda\succeq 0
\end{equation}
This problem is called the \emph{Lagrange dual problem} associated with the problem (\ref{opt}).
The problem (\ref{opt}) is sometimes called
the \emph{primal problem}. We refer to $(\lambda,\nu)$ as
\emph{dual optimal} if they are optimal for the problem
(\ref{dual}).
\par
The optimal value of the Lagrange dual problem, which we denote
$d^{\star}$, is, by definition, the best lower bound on
$p^{\star}$ that can be obtained from the dual function. We have
\begin{equation}
d^{\star} \leqslant p^{\star}
\end{equation}
We refer to the difference $p^{\star}-d^{\star}$ as the
\emph{optimal duality gap} of the original problem (\ref{opt}).
\par
We now assume that functions $f_{0},...,f_{m},h_{1},...,h_{p}$ are
differentiable. Let $x^{\star}$ and
$(\lambda^{\star},\nu^{\star})$ be any primal and dual optimal
points with zero duality gap. Then we have
$$
f_{i}(x^{\star})\leq0,\quad i=1,...,m
$$
$$
h_{i}(x^{\star})=0,\quad i=1,...,p
$$
$$
\lambda_{i}^{\star} \geq 0,\quad i=1,...,m
$$
$$
\nabla f_{0}(x^{\star}) + \sum^{m}_{i=1}\lambda_{i}^{\star} \nabla
f_{i}(x^{\star})+\sum^{p}_{i=1}\nu^{\star}_{i} \nabla
h_{i}(x^{\star})=0
$$
\begin{equation}
\lambda_{i}^{\star}f_{i}(x^{\star})= 0,\quad i=1,...,m
\end{equation}
which are called the Karush-Kuhn-Tucker (KKT) conditions. The
condition $\lambda^{\star}_{i} f_{i}(x^{\star})= 0,\quad
i=1,...,m$ is known as \emph{complementary slackness}; it holds
for any primal optimal $x^{\star}$ and any dual optimal
$(\lambda^{\star},\nu^{\star})$
(when duality gap is zero).
\par
The converse holds, if the primal problem is convex. In other
words, if $\tilde{x}$, $\tilde{\lambda}$, $\tilde{\nu}$ are any
points that satisfy the KKT conditions and $f_{i}$ are convex and
$h_{i}$ are affine, then $\tilde{x}$ and
$(\tilde{\lambda},\tilde{\nu})$ are primal and dual optimal, with
zero duality gap.\\

\vspace{1cm}\setcounter{section}{0} \setcounter{equation}{0}
\renewcommand{\theequation}{B-\roman{equation}}
{\Large{Appendix B}}\\

\textbf{Proof of (\ref{result3})}\\
\par
By taking dot products of (\ref{derivative8}) by $\textbf{c}_{1}$
and $\textbf{c}_{2}$, and then using them and equations
(\ref{derivative9}) and (\ref{derivative1}), the \emph{Lagrange
multipliers} associated with inequality constraints can be
expressed as
\begin{equation}\label{lambda1}
\lambda_{1}=\frac{(1-\tilde{p}_{1})[(\textbf{c}_{1}.\textbf{c}_{2})(\textbf{c}_{2}.\textbf{c}_{3})-(\textbf{c}_{1}.\textbf{c}_{3})]}{2[1+(\textbf{c}_{1}.\textbf{c}_{2})-(\textbf{c}_{1}.\textbf{c}_{3})-(\textbf{c}_{2}.\textbf{c}_{3})][1-(\textbf{c}_{1}.\textbf{c}_{2})]},
\end{equation}
\begin{equation}\label{lambda2}
\lambda_{2}=\frac{(1-\tilde{p}_{2})[(\textbf{c}_{1}.\textbf{c}_{2})(\textbf{c}_{1}.\textbf{c}_{3})-(\textbf{c}_{2}.\textbf{c}_{3})]}{2[1+(\textbf{c}_{1}.\textbf{c}_{2})-(\textbf{c}_{1}.\textbf{c}_{3})-(\textbf{c}_{2}.\textbf{c}_{3})][1-(\textbf{c}_{1}.\textbf{c}_{2})]},
\end{equation}
\begin{equation}\label{lambda3}
\lambda_{3}=\frac{(1-\tilde{p}_{3})[1+(\textbf{c}_{1}.\textbf{c}_{2})]}{2[1+(\textbf{c}_{1}.\textbf{c}_{2})-(\textbf{c}_{1}.\textbf{c}_{3})-(\textbf{c}_{2}.\textbf{c}_{3})]}.
\end{equation}
Taking dot products of (\ref{derivative8}) by $\textbf{c}_{2}$ and
$\textbf{c}_{3}$ and in the same manner, we find
\begin{equation}
\lambda_{1}=\frac{(1-\tilde{p}_{1})[(\textbf{c}_{2}.\textbf{c}_{3})^{2}-1]}{2[2(\textbf{c}_{1}.\textbf{c}_{2})(\textbf{c}_{1}.\textbf{c}_{3})(\textbf{c}_{2}.\textbf{c}_{3})-(\textbf{c}_{1}.\textbf{c}_{3})(\textbf{c}_{2}.\textbf{c}_{3})-(\textbf{c}_{1}.\textbf{c}_{2})(\textbf{c}_{2}.\textbf{c}_{3})-(\textbf{c}_{1}.\textbf{c}_{2})^{2}-(\textbf{c}_{1}.\textbf{c}_{3})^{2}+(\textbf{c}_{1}.\textbf{c}_{2})+(\textbf{c}_{1}.\textbf{c}_{3})]},
\end{equation}
\begin{equation}
\lambda_{2}=\frac{(1-\tilde{p}_{2})[(\textbf{c}_{1}.\textbf{c}_{2})-(\textbf{c}_{1}.\textbf{c}_{3})(\textbf{c}_{2}.\textbf{c}_{3})]}{2[2(\textbf{c}_{1}.\textbf{c}_{2})(\textbf{c}_{1}.\textbf{c}_{3})(\textbf{c}_{2}.\textbf{c}_{3})-(\textbf{c}_{1}.\textbf{c}_{2})(\textbf{c}_{2}.\textbf{c}_{3})-(\textbf{c}_{1}.\textbf{c}_{3})(\textbf{c}_{2}.\textbf{c}_{3})-(\textbf{c}_{1}.\textbf{c}_{2})^{2}-(\textbf{c}_{1}.\textbf{c}_{3})^{2}+(\textbf{c}_{1}.\textbf{c}_{2})+(\textbf{c}_{1}.\textbf{c}_{3})]},
\end{equation}
\begin{equation}
\lambda_{3}=\frac{(1-\tilde{p}_{3})[(\textbf{c}_{1}.\textbf{c}_{3})-(\textbf{c}_{1}.\textbf{c}_{2})(\textbf{c}_{2}.\textbf{c}_{3})]}{2[2(\textbf{c}_{1}.\textbf{c}_{2})(\textbf{c}_{1}.\textbf{c}_{3})(\textbf{c}_{2}.\textbf{c}_{3})-(\textbf{c}_{1}.\textbf{c}_{2})(\textbf{c}_{2}.\textbf{c}_{3})-(\textbf{c}_{1}.\textbf{c}_{3})(\textbf{c}_{2}.\textbf{c}_{3})-(\textbf{c}_{1}.\textbf{c}_{2})^{2}-(\textbf{c}_{1}.\textbf{c}_{3})^{2}+(\textbf{c}_{1}.\textbf{c}_{2})+(\textbf{c}_{1}.\textbf{c}_{3})]}.
\end{equation}
These equations lead to the equation (\ref{result3}).\\
\newpage
\vspace{1cm}\setcounter{section}{0} \setcounter{equation}{0}
\renewcommand{\theequation}{A-\roman{equation}}

\end{document}